\documentstyle[12pt,aasms4]{article}

\begin{document}

\title{Multiphase Chemical Evolution in Elliptical Galaxies}
\author{Yutaka Fujita}
\affil{Graduate School of Human and Environmental Studies, \\
Kyoto University, Kyoto 606-01, Japan}
\affil{Yukawa Institute for Theoretical Physics, \\
Kyoto University, Kyoto 606-01, Japan}
\authoraddr{Kyoto 606-01, Japan}
\authoremail{fujita@yukawa.kyoto-u.ac.jp}
\author{Junji Fukumoto}
\affil{Cray Research Japan, Cuore Bldg., 9th Floor, 12-25, 
Hiroshiba-cho,\\
Suita-si, Osaka 564, Japan}
\authoremail{fukumoto@crj.cray.com}
\author{Katsuya Okoshi}
\affil{Department of Earth and Space Science,Faculty of Science,\\
Osaka University,Machikaneyama-cho, Toyonaka, Osaka 560, Japan}
\authoremail{okoshi@vega.ess.sci.osaka-u.ac.jp}

\begin{abstract}
The recent {\em ASCA} results show that the iron abundance of the
X-ray gas of elliptical galaxies is less than the solar
abundance ($Z_{\sun}$). The observed low iron
abundance is inconsistent with the predictions of the previous
chemical evolution models. In order to solve this problem, 
we present a simple model of chemical evolution for
elliptical galaxies after the galactic wind period 
under the assumption that the gases ejected from
stars do not mix with the circumferential gas.
The ejected gas components 
evolve separately according to their birth time and origin.
We have investigated their evolution qualitatively. 
The gas components originated from supernova
remnant shells cool and drop out of the hot gas 
faster than the other components
because of their high density and metal abundance. As 
a result, supernovae cannot heat the whole gas of the elliptical
galaxies effectively, 
in contrary to the previous results. If the metal abundance of
mass-loss 
gas is not uniform, the mass-loss gas with higher
abundance also
easily drops out and the average abundance can decrease.
We believe that this is a hint of solving the low abundance
problem. 
\end{abstract}


\section{INTRODUCTION}
\label{sec-1}
\indent

It is now well established by X-ray observations that 
elliptical galaxies are often sources of thermal X-ray
emission. Temperatures of the hot interstellar medium (ISM) are
well-determined to be 
around 1 keV and X-ray luminosities are typically around $10^{40}$ erg. 
In contrast, until recently, we have had little
information on the metal abundance and the spatial structure of the
hot ISM due to the poor energy and spatial resolution of detectors.
For this reason, in most of works on the chemical evolution 
for elliptical galaxies, it had been assumed that 
the gas ejected from stars spontaneously mixes with the ISM.  
This inevitably leads to the metal abundance of the 
ISM higher than the value for the sun, because 
the metal abundance of the stars in elliptical galaxies is 
higher than the value for the sun
(e.g. Arimoto \& Yoshii \markcite{ay1987}1987 ; 
Matteucci \& Tornamb\'{e} \markcite{mt1987}1987 ; 
Mihara \& Takahara \markcite{mg1995}1994 ; 
Matteucci \& Gibson \markcite{mg1995}1995 ; 
Fukumoto \& Ikeuchi \markcite{fi1996}1996). 

The recent observations of {\em ASCA} (Awaki et al. 
\markcite{amt1994} 1994 ; Matsushita
et al.\markcite{mma1994} 1994) have changed the
situation of this problem significantly. 
The {\em ASCA} observations showed that the iron  
abundance of the hot ISM is less than that of the sun, 
which is inconsistent with the predictions of the previous
simple chemical evolution models. 

A hint of solving this problem is given by the recent {\em ROSAT}
observation and theoretical argument. First, the recent {\em ROSAT}
observation showed the inhomogeneity and the
existence of cooling clumps in the hot ISM of NGC 507.
(Kim \& Fabbiano
\markcite{kf1995} 1995)
Second, Mathews\markcite{m1990} 
(1990) indicated that
the $\sim 1 M_{\sun}$ of metal ejected by each supernova event into the ISM 
is trapped locally within the hot bubbles and will never
mix thoroughly with the ambient hot gas. He also argued that metal 
ejected through stellar wind may not mix either and 
make small fluctuation of metal 
abundance in the ISM. 
These observational and theoretical arguments suggest that 
the ISM is not well mixed, which is inconsistent with
the assumption of the previous chemical evolution models.
If this is the case, we expect that gas components with higher metal 
abundance
and density will cool faster than other components ; they go into cool 
phase and do not emit X-ray. 
This will reduce effectively the observed average abundance of the hot ISM. 
Renzini et al. \markcite{rcdp1993}(1993) has already explored 
the similar idea based on two-phase model and showed that  
the model does not reproduce the observed iron K line of 
NGC 1399, the cD galaxy in
the Fornax cluster. 
Although the galaxies we investigated here are non-cD
galaxies, 
we suggest that their two-phase model is too simple to
explain the observation. 
In this paper, we investigate 
the chemical evolution of elliptical galaxies assuming that the 
gas released from stars does not mix with the ambient 
ISM ; we propose the multiphase model consisting of hundreds 
of gas phases and consider their evolutions.

The paper is organized as follows. First, in the next section we
describe the fundamental assumptions in detail and write down the
basic equations to determine the evolution of the iron abundance
of the ISM. Then, in \S3, we solve these equations numerically for some
typical models. On the
basis of this general analysis, in \S4, we simulate the X-ray 
spectra of our model galaxies and discuss the possibility of
accounting for the observed low iron abundance of the
hot ISM 
by comparing these simulations with the observational data. 
Our conclusions are summarized in \S5. 

\section{ASSUMPTION AND BASIC EQUATIONS}
\label{sec-2}
\indent


We consider the chemical evolution after galactic wind period.
We assume for simplicity that an elliptical galaxy is a sphere 
of radius $R$ and it is uniform on average, although it contains many
gas components (hereafter called ``phases''). The phases are classified 
by their origin and birth time.
We consider the three types of the origin, that is, 
gas ejected through stellar wind (mass-loss
gas), shell and cavity of supernova remnants (SNRs).
From now on, the phases of each type are called ``mass-loss phase'',
``shell phase'', and ``cavity phase'' indicated by the indices 
$ML$, $s$, and $c$, respectively.
We divide time into a finite number of steps, 
$0<t_{0}<t_{1}<...<t_{i}<...<t_{n}$, where $t_{0}$ and $t_{n}$ 
are the start and end time of the calculation, respectively.
We define $t=t_{0}(=0.5 \rm Gyr)$ 
as the time when the galactic wind stops.
In each time-step,
one phase is born for each type.
The phase of type $\alpha$ ($ML$, $s$, or $c$) 
which is born in $t_{i-1} < t < t_{i}$
is called the $(i,\alpha)$-th phase and 
has its own temperature $T^{(i,\alpha)}$, 
density $\rho^{(i,\alpha)}$, metal abundance $Z^{(i,\alpha)}$, and
mass $M^{(i,\alpha)}$.
Each
phase radiates its thermal energy and evolves. We assume that energy
transfer between the phases is worked only by pressure and ignore
thermal conduction for simplicity.
The phases are assumed to be in pressure equilibrium
because sound crossing time of an elliptical galaxy is far 
shorter than its age (see Eq.(\ref{step})). 
Exceptionally, the phases whose
temperatures become below $T_{\rm crit} (= 10^{5}\rm K)$ are not
considered to be in the pressure equilibrium 
because the condition of the pressure equilibrium will 
broken down for the phases due to the high cooling rate, 
although it also depends on size of each gas blob 
composing the phases (Cowie, Fabian, \& Nulsen\markcite{cfn1980} 
1980). They are assumed to 
cool immediately and drop out of the hot ISM. This means that while the
total number of the type of the phase is unchanged, the number of the 
phase whose temperature is above $T_{\rm crit}$ 
can be different among the types at the given time. 
The gas 
left after galactic wind is also considered as a phase called
``zero-phase''or $(0,0)$-th phase. 

In \S\ref{sec-fm}, we describe the formation of the phases, and 
$t=t_{i}\:(1 \leq i \leq n)$ unless otherwise mentioned. 
In \S\ref{sec-beq}, we describe the 
evolution of the phases and the galaxy.

\subsection{FORMATION OF EACH PHASE}
\label{sec-fm}

\subsubsection{Mass-Loss Phases}
\label{sec-ml}
\indent

Here, we describe the formation of mass-loss phases.
The relation between the mass of a star $m(\tau)$, and its lifetime  
$\tau$, is given by 
\begin{equation}
\label{turn}
\log_{10}m(\tau) = 
1.983 - 1.054\sqrt{\log_{10}\tau + 2.52}
\:,
\end{equation}
where the mass is in units of solar mass ($M_{\sun}$) and the lifetime is in
units of Gyr.
(Larson\markcite{l1974} 1974). 
Because we consider the chemical evolution after galactic wind 
stops ($t> t_{0}=0.5 $Gyr), 
this equation implies that we do not have to consider stars with mass
larger than $2.85 M_{\sun}$ in our
calculation period. 
We assume that
stars with mass in the range of $0.1 - 2.85 M_{\sun}$ 
lose their masses by stellar winds ; 
we simply assume that the mass loss occurs instantaneously 
at the end of the life given by the fraction $f$ of the
initial mass of a star with $m$, 
\begin{equation}
\label{frac-ml}
f = \left\{\begin{array}{ll}
                     0 & \mbox{for $m \leq 0.7$}\:,\\
0.42 m & \mbox{for
$0.7 < m \leq 1.0$}\:,\\ 
0.8 - 0.43/m  & \mbox{for
$1.0 < m \leq 2.85$}\:,\\
 		 \end{array}	 \right. 
\end{equation}
(K\"{o}ppen \& Arimoto\markcite{ka1991} 
1991).
Further, we assume that the temperature of the gas ejected 
by the stellar wind immediately becomes equal to 
the virial temperature of the galaxy $\hat{T}$ for
simplicity.  

Under this assumption, 
the initial temperature and density of the mass-loss 
phase are then given by 
\begin{equation}
T^{(i,ML)}(t_{i}) =  \hat{T} \:,
\end{equation}
\begin{equation}
\rho^{(i,ML)}(t_{i}) = \frac{\mu m_{\rm H} P(t_{i})}{k_{\rm B}
\hat{T}} \:,
\end{equation}
where $\mu$ is the mean molecular weight ($=0.6$), $m_{\rm H}$ is 
the mass of the hydrogen atom, $k_{\rm B}$ is Boltzmann constant, and 
$P(t)$ is the average pressure of 
the ISM. The pressure 
$P(t)$ is obtained by solving the evolution equations of the galaxy 
described in \S\ref{sec-beq}.

The gas and iron 
mass of the mass-loss phase at its birth time,  
$M^{(i,ML)}(t_{i})$  
and $M_{\rm Fe}^{(i,ML)}(t_{i})$, respectively, are given by 
\begin{equation}
M^{(i,ML)}(t_{i})=\int_{t_{i-1}}^{t_{i}}L_{\star}^{(ML)}(t)dt \: ,
\end{equation}
\begin{equation}
M_{\rm Fe}^{(i,ML)}(t_{i})=\int_{t_{i-1}}^{t_{i}}
L_{\rm Fe}^{(ML)}(t)dt \:,
\end{equation}
where 
$L_{\star}^{(ML)}$ and $L_{\rm Fe}^{(ML)}$ 
are the gas and iron 
mass ejection rates from stars, respectively. 
Since the time scale of star formation in an elliptical galaxy, which is
typically $10^{7-8}$ yr, is short enough 
compared with the galaxy age 
(e.g. Arimoto \& Yoshii\markcite{ay1987} 1987), we assume
that the stellar system of the galaxy formed at $t=0$ simultaneously. 
Thus, $L_{\star}^{(ML)}$ and $L_{\rm Fe}^{(ML)}$ are given by 
\begin{equation}
\label{star-ml}
L_{\star}^{(ML)}(t) = f \left|\frac{d m(\tau)}{d\tau}
\right|_{\tau=t} \phi(m(\tau=t)) M_{\star}(0)  \:,
\end{equation}
\begin{equation}
\label{fe-ml}
L_{\rm Fe}^{(ML)}(t) = Z_{\rm ML} f \left|
\frac{d m(\tau)}{d\tau}\right|_{\tau=t}
\phi(m(\tau=t))  M_{\star}(0)  \:,
\end{equation}
where $\phi(m)$, $M_{\star}(t)$ and $Z_{\rm ML}$ are the initial mass
function (IMF), the stellar mass of the galaxy and 
the iron abundance of the mass-loss gas, respectively ; 
the power of the IMF is taken to be 1.35 and the other variables
are given by later.

The density $\rho^{(i,ML)}(t_{i})$, the temperature 
$T^{(i,ML)}(t_{i})$, the mass $M^{(i,ML)}(t_{i})$, and the iron mass $M_{\rm
Fe}^{(i,ML)}(t_{i})$ determined by above equations give 
the initial conditions of the
evolution equations of the phases (see \S\ref{sec-beq}).

\subsubsection{Shell and Cavity Phases}
\label{sec-sn}
\indent

Next, we describe the formation of the shell and cavity phases. 
Since the progenitors of Type II supernovae have short lives 
$\tau<t_{0}$, Type II
supernovae are important only in the galactic wind period 
(e.g. Arimoto \& Yoshii\markcite{ay1987} 1987 ; Matteucci \& 
Tornamb\'{e}\markcite{mt1987} 1987 ; David, Forman, \& Jones
\markcite{dfj1991} 1991 ; Matteucci \& Gibson\markcite{mg1995} 1995).
Therefore, we consider only Type Ia supernovae (SN Ia) 
because we are concerned with chemical
evolution after galactic wind stops. 

We use the adiabatic Sedov solution in pressure equilibrium 
to determine the initial conditions of the shell and cavity phases for 
simplicity.
The actual shape of the SNRs is deviated from the
Sedov solution because of the high temperature of the
unshocked gas. However, we confirmed that 
the deviation is not important 
(see \S\ref{sec-anl}).

For simplicity, we assume that the SNRs are consists of two parts,
outer shell and inner cavity region, and
that they evolve as separate phases. 
We call the former ``shell phase'' and the latter ``cavity
phase'', denoted by the indices $s$ and $c$, respectively.
When the distance from the center of the SNR is $r$ and 
the shock front radius is $r_{\rm s}$,   
the inner cavity region and the outer shell region correspond to 
$r<(1-k)r_{\rm s}$ and $(1-k)r_{\rm s} < r < r_{\rm s}$, respectively, 
where $k$ is representing the width of the shell.  
In each region, the average density and temperature
are decided as follows. 

The shock front radius is given by 
\begin{equation}
\label{sn-front}
r_{\rm s}=\left(\frac{2.02 E_{\rm SN}t_{\rm s}^{2}}{\rho}\right)^{1/5}
\end{equation}
(Spitzer\markcite{s1978} 1978), where $E_{SN}(=10^{51}$erg) 
is the energy released from 
a supernova, 
$t_{s}$ is the time past since the supernova explosion,  
and $\rho$ is the average density of the ISM.
The pressure just inside the shock front is given by 
\begin{equation}
P_{1}=\frac{2\rho}{(\gamma+1)}
      \left(\frac{dr_{\rm s}}{dt_{\rm s}}\right)^{2}
     =\frac{8\rho^{3/5}(2.02 E_{\rm SN})^{2/5}t_{s}^{-6/5}}{25(\gamma+1)}
\end{equation}
(Spitzer\markcite{s1978} 1978), 
where $\gamma (=5/3)$ is the adiabatic constant.

We assume that $P_{1}$ is equal to the average pressure of the ISM
$P$. Under the assumption,
the time $t_{\rm s}$ is given by 
\begin{equation}
\label{eq-time}
t_{s}=\left[\frac{8}{25(\gamma+1)}\right]^{5/6} 
      P^{-5/6} (2.02 E_{\rm SN})^{1/3}\rho^{1/2}  \:, 
\end{equation}
and
\begin{equation}
\label{rs}
r_{\rm s} = \left[\frac{8}{25(\gamma+1)}\right]^{1/3}
         P^{-1/3} (2.02 E_{\rm SN})^{1/3} \:.
\end{equation}
The density and pressure distribution of the Sedov solution,  
$\rho_{\rm Sedov}(r)$ and $P_{\rm Sedov}(r)$, respectively, 
are determined uniquely under the condition $P=P_{1}$.

In terms of $r_{s}$ and $\rho_{\rm Sedov}(r)$,  
the average density in the
shell region is expressed by 
\begin{equation}
\label{eq-rhos}
\rho_{s}=\int_{(1-k)r_{\rm s}}^{r_{\rm s}}4\pi \rho_{\rm Sedov}(r) r^{2}dr/
         \int_{(1-k)r_{\rm s}}^{r_{\rm s}}4\pi r^{2}dr \:.
\end{equation}
The mass inside the shock front is given by 
\begin{equation}
m_{\rm SNR}=\frac{4}{3}\pi r_{\rm s}^{3}\rho + m_{\rm pro} \:,
\end{equation}
where $m_{\rm pro}$ is the mass of the progenitor star. 
The mass in the shell region is given by 
\begin{equation}
m_{\rm s}=\rho_{\rm s}\int_{(1-k)r_{\rm s}}^{r_{\rm s}}4\pi r^{2}dr \:.
\end{equation}
Hence, the mass and average density in the cavity region are given by 
\begin{equation}
m_{\rm c}=m_{\rm SNR}-m_{\rm s} \:,
\end{equation}
\begin{equation}
\rho_{\rm c}= \frac{3m_{\rm c}}{4\pi (1-k)^{3} r_{\rm s}^{3}} \:.
\end{equation}

The iron abundance 
is assumed to be uniform inside the shock front radius  
$r_{\rm s}$, 
because Mathews \markcite{m1990}(1990) estimated that 
in an elliptical galaxy iron fragments
made by Rayleigh -  Taylor instability reach the shock front
$\sim 10^{4}$ yr after the supernova
explosion.
The iron mass in the cavity and shell region are 
\begin{equation}
m_{\rm Fe,s}=m_{\rm c}\left(\frac{m_{\rm Fe}}{m_{\rm SNR}}+\hat{Z}
\right)\:, 
\end{equation}
\begin{equation}
m_{\rm Fe,c}=m_{\rm s}\left(\frac{m_{\rm Fe}}{m_{\rm SNR}}+\hat{Z}
\right)\:, 
\end{equation}
where $m_{\rm Fe}$ is the iron mass ejected by the supernova and
$\hat{Z}$ is the average iron abundance of the ISM occupied by the SNR.

Next, we determine the temperatures. For the Sedov solution, 
the specific thermal energy 
in the shell region is given by 
\begin{equation}
\label{eq-us}
U_{s}=\frac{1}{m_{\rm s}}\int_{(1-k)r_{\rm s}}^{r_{\rm s}}
      \frac{P_{\rm Sedov}}{\gamma-1}4\pi r^{2}dr \:.
\end{equation}
The specific thermal energy 
in the cavity region is given by 
\begin{equation}
U_{\rm c}=(E_{\rm SN}-m_{\rm s}U_{\rm s})/m_{\rm c} \:.
\end{equation}
Since the thermal energy of the ISM occupied by the SNR cannot be
ignored, the average temperatures of the shell and 
cavity region are given by 
\begin{equation}
T_{\rm s}=\frac{2}{3}\frac{\mu m_{\rm H} U_{\rm s}}{k_{\rm B}}
+\hat{T}\:,
\end{equation}
\begin{equation}
\label{tc}
T_{\rm c}=\frac{2}{3}\frac{\mu m_{\rm H} U_{\rm c}}{k_{\rm B}}
+\hat{T}\:,
\end{equation}
respectively, where $\hat{T}(=P\mu m_{\rm H}/k_{\rm B}\rho)$ 
is the average temperature 
of the ISM which is assumed to be 
equals to the virial temperature. 
The densities and temperatures determined so far are adopted as
the initial values of the
shell and cavity phase for the evolution equations of the phases 
described in \S\ref{sec-beq} as follows : 
\begin{equation}
\rho^{(i,s)}(t_{i})=\rho_{s} \:,
\end{equation}
\begin{equation}
T^{(i,s)}(t_{i})=T_{s} \:,
\end{equation}
\begin{equation}
\rho^{(i,c)}(t_{i})=\rho_{c} \:, 
\end{equation}
\begin{equation}
T^{(i,c)}(t_{i})=T_{c} \:.
\end{equation}
From Eqs. (\ref{rs}) - (\ref{tc}), they are determined uniquely if 
$E_{\rm SN}$, $m_{\rm pro}$, $m_{\rm Fe}$, $k$, $\rho$, $\hat{T}$, and
$\hat{Z}$ are given. We fix $E_{\rm SN}$,
$m_{\rm pro}$, and $m_{\rm Fe}$ by giving typical values, that is, 
$m_{\rm pro}=2.0 M_{\sun}$, $m_{\rm Fe}=0.75\rm M_{\sun}$,  
and $E_{\rm SN}=10^{51}$
erg. We also fix $k=0.3$ ; the results in \S\ref{sec-anl} are 
qualitatively unchanged even if we take the
different value for $0.25 \lesssim k \lesssim 0.45$. 
The other 
parameters, $\rho$, $\hat{T}$, and $\hat{Z}$ are time-dependent and 
determined by solving the evolution
equations of the galaxy described in \S\ref{sec-beq}. 
For examples, for $\rho = 1.67 \times
10^{-27} \rm g \; cm^{-3}$, $\hat{T}=1.3$ keV, and $\hat{Z}=1.0 \rm 
Z_{\sun}$, we can estimate each quantity : 
$\rho_{s}=2.4 \times 10^{-27} \rm g \; cm^{-3}$,
$\rho_{c}=2.3 \times 10^{-28} \rm g \; cm^{-3}$, $T_{s}=1.8$ keV, 
$T_{c}=9.6$ keV, and $m_{\rm
Fe,s}/m_{s}=m_{\rm Fe,c}/m_{c}=2.8 \rm Z_{\sun}$. 
Note that since the cooling function is nearly independent of temperature
and inversely proportional to metal abundance near $T \sim 10^{7} \rm K$ 
(see Eq.(\ref{eq-cool})), the cooling time is proportional to 
$T/(\rho Z)$. Thus, 
the cooling time of the shell phase ($\propto T_{s}/(\rho_{s}m_{\rm
Fe,s}/m_{s})$) is generally shorter than that of the ambient medium ($\propto 
\hat{T}/(\rho \hat{Z})$). It causes selective 
cooling of the shell phase. This effect will be discussed in
\S\ref{sec-anl}. 

The gas and iron 
mass of the shell and cavity phase at their birth time 
are given by 
\begin{equation}
M^{(i,s)}(t_{i})=\int_{t_{i-1}}^{t_{i}}L_{\star}^{(s)}(t)dt \: ,
\end{equation}
\begin{equation}
M^{(i,c)}(t_{i})=\int_{t_{i-1}}^{t_{i}}L_{\star}^{(c)}(t)dt \: ,
\end{equation}
\begin{equation}
M_{\rm Fe}^{(i,s)}(t_{i})=\int_{t_{i-1}}^{t_{i}} 
L_{\rm Fe}^{(s)}(t)dt \:,
\end{equation}
\begin{equation}
M_{\rm Fe}^{(i,c)}(t_{i})=\int_{t_{i-1}}^{t_{i}} 
L_{\rm Fe}^{(c)}(t)dt \:,
\end{equation}
where 
$L_{\star}^{(s)}$, $L_{\star}^{(c)}$, $L_{\rm Fe}^{(s)}$, and 
$L_{\rm Fe}^{(c)}$ 
are the mass and iron production rates of the shell and cavity phase,
respectively ; they are given by 
\begin{equation}
\label{star-s}
L_{\star}^{(s)}(t) = r_{\rm SN}(t) m_{\rm s}(t) \:,
\end{equation}
\begin{equation}
\label{star-c}
L_{\star}^{(c)}(t) = r_{\rm SN}(t) m_{\rm c}(t) \:,
\end{equation}
\begin{equation}
\label{fe-s}
L_{\rm Fe}^{(s)}(t) = r_{\rm SN}(t) m_{\rm Fe,s}(t) \:,
\end{equation}
\begin{equation}
\label{fe-c}
L_{\rm Fe}^{(c)}(t) = r_{\rm SN}(t) m_{\rm Fe,c}(t) \:,
\end{equation}
where $r_{\rm SN}(t)$ is the SN Ia rate.
The time dependence of the SN Ia rate is
assumed to be $r_{\rm SN}(t) \propto t^{-0.5}$ (David, Forman, \& Jones 
\markcite{dfj1990} 1990).
The normalization is given by the SN Ia rate at $t=t_{n}$,
which is specified later. 

\subsection{EVOLUTION OF THE PHASES AND THE GALAXY}
\label{sec-beq}
\indent

The energy equation for the $(i,\alpha)$-th phase for $t > t_{i}$ is  
given by 
\begin{equation}
\label{energy}
\frac{\rho^{(i,\alpha)}(t)}{\gamma-1}\frac{d}{dt}\left(\frac{k_{\rm B}
T^{(i,\alpha)}(t)}{\mu
m_{\rm H}}\right)
-\frac{k_{\rm B}T^{(i,\alpha)}(t)}
{\mu m_{\rm H}}\frac{d}{dt}\rho^{(i,\alpha)}(t)
=-(n_{e}^{(i,\alpha)})^{2}
\Lambda(Z^{(i,\alpha)},T^{(i,\alpha)}(t)) \:,
\end{equation}
where 
$n_{e}^{(i,\alpha)}$ is the electron density 
of the $(i,\alpha)$-th phase and $\Lambda$ 
is the cooling function. 
Assuming that the relative metal abundance is the same as that of the 
sun for simplicity,
the cooling function is approximated by
\begin{eqnarray}
\label{eq-cool}
\Lambda(Z^{(i,\alpha)}, T^{(i,\alpha)}) 
 & = &\left[2.1 \times 10^{-27}
      \left(1 + 0.1\frac{Z^{(i,\alpha)}}{\rm Z_{\sun}}\right) 
      \left(\frac{T^{(i,\alpha)}}{\rm K} \right)^{0.5}\right.
      \nonumber \\
 & + &\left.\left(0.04 + \frac{Z^{(i,\alpha)}}{\rm Z_{\sun}}\right)
      6.2 \times 10^{-19}
      \left(\frac{T^{(i,\alpha)}}{\rm K}\right)^{-0.6}\right] 
      \nonumber \\
 &   & (\rm ergs \: cm^{-3} s^{-1}) \:.
\end{eqnarray}
This is an empirical formula derived by fitting to the cooling curves in
Figure 9-9 of Binney \& Tremaine \markcite{bt1987}(1987).

In our model, all existing phases are assumed to be in pressure
equilibrium because sound crossing time of an elliptical galaxy is far 
shorter than its age (see Eq.(\ref{step})). Thus, for $i<j$, 
\begin{equation}
\label{peq}
\rho(t_{j}) \frac{k_{\rm B}\hat{T}}{\mu m_{\rm H}}
 =\rho^{(i,\alpha)}(t_{j})\frac{k_{\rm B}T^{(i,\alpha)}(t_{j})}{\mu m_{\rm H}}
 = P(t_{j}) \:,
\end{equation}
\begin{equation}
\label{rho}
\rho(t_{j})=\frac{M_{\rm g}(t_{j})}{V(t_{j})} \:,
\end{equation}
where $M_{\rm g}$ and $V$ are the total gas mass and volume of the
galaxy, respectively.
Here, we have approximated $dP/dR$ by $P/R$ for simplicity. 
Note that $V=(4\pi/3) R^{3}$.

The total gas mass, the iron mass and the gas volume are the 
summation of those of each phase ; 
\begin{equation}
\label{mg}
M_{\rm g}(t)=\sum_{\alpha}
             \sum_{i,\: T^{(i,\alpha)}>T_{\rm crit}}M^{(i,\alpha)}(t) ,
\end{equation}
\begin{equation}
\label{mfe}
M_{\rm Fe}(t)=\sum_{\alpha}
          \sum_{i,\: T^{(i,\alpha)}>T_{\rm crit}}M_{\rm Fe}^{(i,\alpha)}(t) ,
\end{equation}
\begin{equation}
\label{volume}
V(t)=\sum_{\alpha}\sum_{i,\: T^{(i,\alpha)}>T_{\rm crit}}
         V^{(i,\alpha)}(t) ,
\end{equation}
where 
\begin{equation}
V^{(i,\alpha)}(t) = 
M^{(i,\alpha)}(t)/\rho^{(i,\alpha)}(t) \:. 
\end{equation}
As mentioned above, the phases whose
temperatures are below $T_{\rm crit} (= 10^{5}\rm K)$ are not included 
in the summation.

Note that part of the SNRs is composed of pre-existing phases. 
As mentioned in \S\ref{sec-sn}, we assumed that the
SNRs evolve as new phases, then the masses of the pre-existing phases
are reduced by the occupation by the SNRs.
Thus, the mass and iron mass 
of the pre-existing $(i,\alpha)$-th phase at $t=t_{j}$ $(i<j)$ are given by 
\begin{equation}
\label{red}
M^{(i,\alpha)}(t_{j}) = M^{(i,\alpha)}(t_{j-1})-
\left. M_{\rm SNR}^{(j)}
\frac{V^{(i,\alpha)}(t)}{V(t)} 
\right|_{t=t_{j-1}} \:,
\end{equation}
\begin{equation}
M_{\rm Fe}^{(i,\alpha)}(t_{j}) = M_{\rm Fe}^{(i,\alpha)}(t_{j-1})-
\left. M_{\rm SNR}^{(j)}
\frac{M_{\rm Fe}^{(i,\alpha)}(t)}{M^{(i,\alpha)}(t)}
\frac{V^{(i,\alpha)}(t)}{V(t)} 
\right|_{t=t_{j-1}} \:,
\end{equation}
respectively, where $M_{\rm SNR}^{(j)}$ is the mass occupied by SNRs during
$t_{j-1}<t<t_{j}$, which is given by
\begin{equation}
M_{\rm SNR}^{(j)}=\int_{t_{j-1}}^{t_{j}}
\frac{4}{3}\pi \rho(t) r_{s}(t)^{3} 
r_{\rm SN}(t) dt \:.
\end{equation}

We take the sound crossing time as the time-steps for the calculations, 
\begin{equation}
\label{step}
t_{j+1}-t_{j}= \frac{R(t_{j})}{\sqrt{G M(R(t_{j}))/R(t_{j})}} 
             = 4.7 \times 10^{7} {\rm yr} 
              \left (\frac{R}{10{\rm kpc}}\right)^{3/2} 
              \left (\frac{M(R)}{10^{11}{\rm M_{\sun}}}\right)^{-1/2} \:.
\end{equation}
Strictly speaking, calculations under Eq.(\ref{step}) are not exact
because the phases whose cooling time is shorter than
the time-step are not treated correctly ;  
these phases are removed at the next time-step $t_{j+1}$ 
and their $PdV$ work in $t_{j} < t < t_{j+1}$ is not taken into
account. 
However, we confirmed that in our examined models 
their $PdV$ work is less than 10\% of the variation 
of the total thermal energy of the galaxy in $t_{j} < t <
t_{j+1}$. Then, we believe that it is negligible for our qualitative 
discussion.

We derive $\rho^{(i,\alpha)}(t_{j})$ and $T^{(i,\alpha)}(t_{j})$ from
$\rho^{(i,\alpha)}(t_{j-1})$ and $T^{(i,\alpha)}(t_{j-1})$ by iterating 
Eqs.(\ref{energy}),
(\ref{peq}), (\ref{rho}), (\ref{mg}), (\ref{volume}), and 
(\ref{red})
until they converge.

\section{NUMERICAL ANALYSIS}
\label{sec-anl}
\indent

The main purpose
of this section is to look into 
the qualitative features of the
multiphase model of the hot ISM of the elliptical galaxy.
We reduce the number of the free parameters by
giving typical fixed values to some of them.
First, for the parameters regarding the whole galaxy, 
we take $t_{0}=0.5$ Gyr, $t_{\rm n}=10$ Gyr, $R(t_{0})=50$ kpc,
$M_{\star}(0)=2\times 10^{11} M_{\sun}$, 
and $M_{\rm g,Fe}(t_{0})/M_{\rm
g}(t_{0})= 4 \rm Z_{\sun}$. We leave $M_{\rm g}(t_{0})$ as a free
parameter. 
Then if we specify $M_{\rm g}(t_{0})$, we can write 
$V(t_{0})^{(0,0)}=(4\pi/3)R(t_{0})^{3}$, $M^{(0,0)}(t_{0})=M_{\rm
g}(t_{0})$, and $M_{\rm Fe}^{(0,0)}(t_{0})=M_{\rm Fe,g}(t_{0})$. 
We take the temperature of the zero-phase 
$T^{(0,0)}(t_{0})=1.3$ keV.
The virial temperature of the galaxy is given by  
\begin{equation}
\hat{T}=T^{(0,0)}(t_{0})\:(=1.3\rm keV)\:.
\end{equation}

Here, we discuss the validity of adopting the Sedov
solution as the initial conditions of the shell and cavity phases 
by comparison with the numerical simulation 
of evolution of a SNR in an elliptical galaxy.
Mathews \markcite{m1990}(1990) calculated the evolution 
of a SNR expanding into ISM with temperature $\hat{T}=10^{7}$ 
K and density $\hat{n}=10^{-3}\rm cm^{-3}$ until the pressure 
equilibrium is achieved. 
Figures 3 and 4 in his paper 
depict temperature and pressure distributions when 
the pressure equilibrium is almost 
achieved ($t_{s}=1.17\times 10^{5}$ yr).
From now on, we call these the ``Mathews solutions''.
Although Mathews showed the variations of temperature and pressure 
with radius only
in the case of $\hat{T}=10^{7}$ K and $\hat{n}=10^{-3}\rm
cm^{-3}$, the fact that the radius 
where the gas temperature has risen by $\sim 2$ 
above the ambient value ($\hat{T}=10^{7}$ K) 
varies as $\hat{n}^{-1/3}$ (his Table 5)
may indicate that the solution is self-similar ; 
we assume the self-similarity in the following argument. 
In the Mathews solution, the 
temperature and pressure gradients of the SNR are somewhat smaller than 
those of the Sedov solution. Thus, to obtain the similar values of
$\rho_{s}$, $\rho_{c}$, $T_{s}$, and $T_{c}$, 
the shell width $k$ must be larger when the 
Mathews solution is used in Eqs. (\ref{eq-rhos}) and
(\ref{eq-us}) than that when the Sedov solution
is used. For example, when $\hat{T}=10^{7}$ 
K and $\hat{n}=10^{-3}\rm cm^{-3}$, the values of $\rho_{s}$, $\rho_{c}$,
$T_{s}$ and $T_{c}$ derived by using the Mathews solution 
and taking $k=0.35$ are equal within 30\% errors 
to those derived by using the Sedov
solution and taking $k=0.3$.
We confirmed that the variation of 30\% does not affect the results in 
this section significantly. Moreover, we confirmed that the results in 
this section are not significantly changed even if we use the Mathews
solution and take $k=0.3$.

We solve the basic 
equations described 
in \S \ref{sec-2} for the models whose details
are given in Table 1. 
The difference between the models A and B is the initial gas mass 
$M_{\rm g}(t_{0})$, which is chosen to match the observed X-ray
luminosities of elliptical galaxies ($\sim 10^{40}\rm erg\: s^{-1}$) 
at $t=t_{n}$ (Table 2). 
Models A1 - A3 and B1 - B3 
are calculated to see the effect of 
SN Ia. 
The SN Ia rate is normalized by the values at $t=t_{n}(=10\rm Gyr)$ 
shown in the third column of Table 1. The values are expressed in units 
of SNu, that is, the number of 
SNe per $10^{10} h^{-2} \rm L_{B\sun}$ per 100 yr ($H_{0}=100h \rm km 
\: s^{-1} \: Mpc^{-1}$ ; we set h=0.5). The ratio of the mass to
luminosity is taken to be
$8M_{\sun}/L_{\sun}$, and the rate for SN Ia is normalized by using
this ratio.
Models A4 - A5 and B4 - B5 are calculated to see the effect of 
metal-abundance distribution for the mass-loss gas.
In these models, the mass-loss phase born in
each time-step is divided into two phases with equal masses and 
different abundances corresponding to the two figures for $Z_{\rm ML}$ 
(Table 1).
On the other hand, in models A1 - A3 and B1 - B3, 
each component has the same
abundance or the mass-loss gas has one kind of the abundance. 
Table 1 also shows total number of the time-step $n$ and the
numbers of phases which have survived 
(their temperature is above
$T_{\rm crit}$) until $t=t_{n}$. 
In models B1 - B3, all phases cool and vanish by sometime before 
$t=t_{n}$, and we
stopped the calculations at that time.
We give the numbers of the survived phases in columns 
$N_{\rm ML,H}$, $N_{\rm ML,L}$, 
$N_{\rm sh}$, and $N_{\rm cav}$
denoting respectively the mass-loss phase with the high and low abundance, 
shell phase, and cavity phase (Table 1). 
Figures 1 - 4 show the time evolution of $R$, $M_{\rm g}$, and
iron abundance $Z$, where $Z$ represents $M_{\rm g,Fe}/M_{\rm g}$ in
units of the  
solar abundance $\rm Z_{\sun}$, $1.7 \times 10^{-3}$.

Figures 1 and 2 show the influence of SN Ia on the gas 
evolution. 
The sharp decreases at $t\sim 4$ Gyr in Figure 1 and at $t\sim 3$ Gyr in
Figure 2 represent cooling and vanishing of the zero-phase. 
Since the initial volume in series A is the same as in series B, 
the initial gas density is higher in series 
B than that in series A. Therefore, the cooling process 
is more effective and $R$
decreases faster in series B than in series A. 
The radius of the gas sphere does not significantly increase 
even if the SN Ia rate at $t=t_{n}$ increases (Fig.1a and 2a). 
Furthermore, the gas mass decrease faster in the models of higher SN Ia
rate (Fig.1b and 2b). 
These facts indicate that supernovae are not effective heating 
sources of the ISM as a whole. 
The reason is that the shell phases cool and drop out 
faster than other phases owing to their high density and metal abundance
(see \S\ref{sec-sn}), and that they have large thermal energies 
when they are born as the following arguments shows. 
When a SNR becomes in pressure equilibrium, the thermal energy within a 
shell region is given by
\begin{equation}
E_{s}=    \frac{m_{s}T_{s}}{m_{s}T_{s}+m_{c}T_{c}}
          \left(E_{\rm SN}+\frac{2\pi k\hat{T}}{\mu m_{\rm H}}
          r_{s}^{3} \rho \right)
\:. 
\end{equation}
Since $E_{s} \gtrsim E_{\rm SN}$ in our parameter range, 
more energy than 
that supplied by supernova explosion is 
radiated from the shell region in a short time. Thus, supernovae 
cannot heat the ISM. 
This is contrary to the previous results (e.g. Ciotti et al. 
\markcite{cdp1991} 1991 ; 
David et al.\markcite{dfj1991} 1991) which claimed that
supernovae are effective as heating sources and swell the ISM.
However, {\em this fact does not reject the galactic wind model.} 
While the expanding SNRs do not overlap in our calculations, 
it is expected that they overlap during the galaxy-formation 
period because of the high supernova rate ; then,    
the ISM is expelled and galactic wind occurs (Ikeuchi 
\markcite{i1977}1977).
Note that 
Mathews \markcite{m1990}(1990) indicated the possibility of buoyant
mixing of the hot cavity region in the local ISM. 
This may decrease entropy fluctuation and the effect of selective
cooling of the shell phases.

In models A1, A2, B1, and B2,
$Z$ is around $\rm Z_{\sun}$ for $t \gtrsim 4$
Gyr (Fig.1c and 2c), because most of the mass of the remaining  
gas is dominated by mass-loss phases ($Z= 1 \rm Z_{\sun}$). 
In models A3 and B3, the shell phases are dominant in mass ; they
start to vanish for $t \gtrsim 6$ in model A3, and for $t \gtrsim 4$ in 
model B3. 

As seen in Figures 3(c) and 4(c), $Z$ is small when the
mass-loss gas has two kinds of the iron abundance. This is because 
the mass-loss phases with the higher iron abundance 
also drop out in addition to the shell phases, as shown in Table 1.
In models A4, A5, B4, and B5, $Z< \rm Z_{\sun}$ at $t=10^{10}$ yr,
which is consistent with the observations (Awaki et al. 
\markcite{amt1994} 1994 ;
Matsushita et al.\markcite{mma1994} 1994). Although our
model may be too simple to give quantitative predictions, 
we believe this is a hint 
of solving the low abundance problem of elliptical galaxies. 

\section{SPECTRAL SIMULATIONS}
\label{sec-sp}
\indent

In most of observations, 
X-ray spectra of galaxies are fitted to at most 
two thin thermal plasma models. However, if the actual ISM is 
composed of many components with various temperatures and metal
abundances, this simple fitting may give deferent values from the 
actual ones. 
In this section, we examine this possibility by constructing 
composite spectra from the many phases which
have survived at $t=t_{n}$(=10 Gyr) and fit them to a few spectral
models by the XSPEC package (ver. 8.50) 
as most of observers did, although our model may be too 
simple for detailed discussion. 

We examine models A1, A2, and B4 because in the other models 
X-ray flux is below the observational limit (Table 2).
The distance to the model galaxies is assumed to be 15 Mpc. 
In our models, several hundred phases are
survived until the present epoch. 
However, the XSPEC package cannot treat so many phases ; for each type 
we collect each 10 phases into a new single phase.
We regard the 
luminosity-weighted temperature and abundance of the 10 phases as 
the temperature and abundance of the new single phase. 
We confirmed that even if we collect each 20 phases into a new single
phase, the results do not change significantly. 
First, the spectrum of each new phase is 
calculated by a thin thermal plasma model (so-called Meka model ; 
Mewe, Gronenschild, and van den Oord\markcite{mgv1985} 
1985 ; Mewe, Lemen, and van den
Oord\markcite{mlv1986} 1986 ; Kaastra\markcite{k1992} 
1992) in the XSPEC package. The reason of using 
Meka model is that the XSPEC package can treat high-abundance
components only with this model. 
Then, using the XSPEC package, the spectral 
components are composed and the Galactic absorption
($10^{21} \rm cm^{-2}$) is taken into account. Since the
cosmic iron abundance to which we referred as the solar abundance is
$2.6 \times 10^{-3}$ in the XSPEC package, 
we corrected it to $1.7 \times 10^{-3}$ 
at composition. 
The model spectra are fitted to 
the Meka model plus an absorption 
with a response function of {\em ASCA} SIS. 
Background is not considered.

The exposure time and the results are shown in Table 2 ;  
the fluxes do not include absorption. 
For models
A1, A2, and B4, the upper lines are the results of the 
fitting to one Meka model and the lower
lines are those to two Meka models. For
comparison, $Z$ at $t=t_{n} (=10$ Gyr) is also shown. 
The abundance $Z_{\rm ML}$, temperature $T_{\rm fit}$, 
and absorption column density $N_{\rm H,fit}$ 
derived by the fitting are also shown in Table 2.
As clearly seen in Table 2, 
$Z_{\rm fit}$ is much smaller than $Z$ when the model spectra are  
fitted to one Meka model plus absorption. However $\chi^{2}$ is so 
large that these models will be rejected ; the fits suffer from significant
positive residuals above $\sim 2$ keV as exemplified in Figure 5. 
We next fitted the spectra with a sum of two Meka models with
different temperatures and normalizations but using common abundance as
Matsushita et al.\markcite{mma1994} (1994) did. 
In these cases, $\chi^{2}$ is significantly reduced. 
The fitted spectra are shown in Figure 6. 
All the models examined here are acceptable and the low abundance, 
$Z_{\rm fit}< Z_{\sun}$, is consistent with the observations (Awaki et
al. 1994 ; Matsushita et al. 1994).
The derived abundance $Z_{\rm fit}$ is still smaller than
$Z$. This may indicate that not only $\chi^{2}$ statics but also 
estimation of equivalent width of the lines is required 
to derive the metal abundance of multiphase gas (Matsushita, private
communication). 
The iron K line is very weak (Figures 5 and 6), because 
of the small number of the survived shell phases (Table 2), and of the 
low density
and high temperature ($\sim 10$ keV) of the cavity phases. The weak 
iron K line is consistent with the observations (Awaki et
al. 1994 ; Matsushita et al. 1994).

\section{SUMMARY AND CONCLUSIONS}
\indent

In this paper, we have presented a model of the chemical evolution for
elliptical galaxies after galactic wind period 
under the assumption that the gases ejected from
stars do not mix with the circumferential gas. The ejected gases 
evolve separately according to their birth time and origin. We
considered three origin of the ejected gas, that is, shell and cavity
of supernova remnants and mass-loss. We fitted the model spectra
made from the survived phases by thin thermal plasma models. 

The main results and conclusions can be summarized as follows:

(i) The model predicts that the supernovae are not effective as heating
sources of the ISM after the galactic wind stops. 
The shells of the supernova remnant can cool and drop out 
rapidly because of
their high density and metal abundance. 
Since the shells initially have large thermal energy, 
the energy ejected by supernova
explosions is radiated and does not heat up the ISM. 

(ii) The present iron abundance of the hot ISM can be less
than $ 1 \rm 
Z_{\sun}$ and consistent with the {\em ASCA} observations if the
abundance of the mass-loss gas has wide distribution. In this case, 
the mass-loss gas with higher abundance also drops out 
in addition to the shells of
the supernova remnant. 
This effect makes the observed average abundance of the ISM 
lower. Although our
model may be too simple to give quantitative predictions, 
we believe this is a hint 
of solving the low abundance problem of elliptical galaxies. 

(iii) If the model spectra are fitted by one thermal spectrum
model, the resultant abundance is well below $ 1 \rm Z_{\sun}$,
although the fits suffer from significant
positive residuals above $\sim 2$ keV. When they are fitted by two
thermal spectrum models, $\chi^{2}$ is significantly reduced and
the derived abundances are also subsolar.

\acknowledgments

We thank N. Gouda, H. Kodama, S. Ikeuchi, N. Arimoto, H. Sato, 
J. Yokoyamka, Y. Yamada, R. Nishi, and T. Shigeyama for useful discussion and 
comments. Y. Fujita thanks K. Koyama for the use of a computer. 
We are also grateful to K. Matsushita and H. Matsumoto for providing
{\em ASCA} data. This work was supported in part by the JSPS Research
Fellowship for Young Scientists.

\newpage

\thispagestyle{empty}

\section*{Figure Captions}

\figcaption[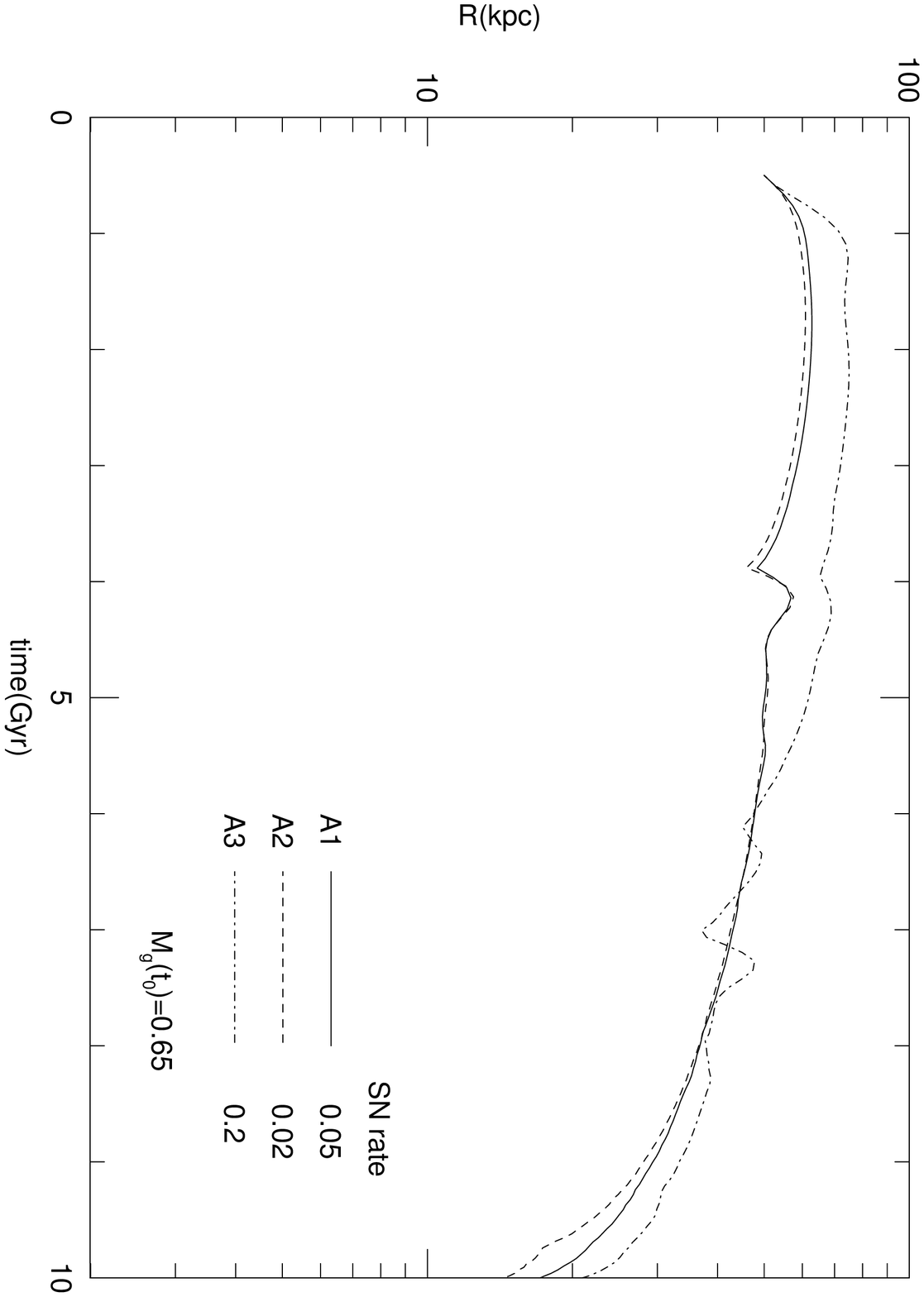]{The evolution of 
(a) radius (b) gas mass and (c) abundance for models A1 - A3.}

\figcaption[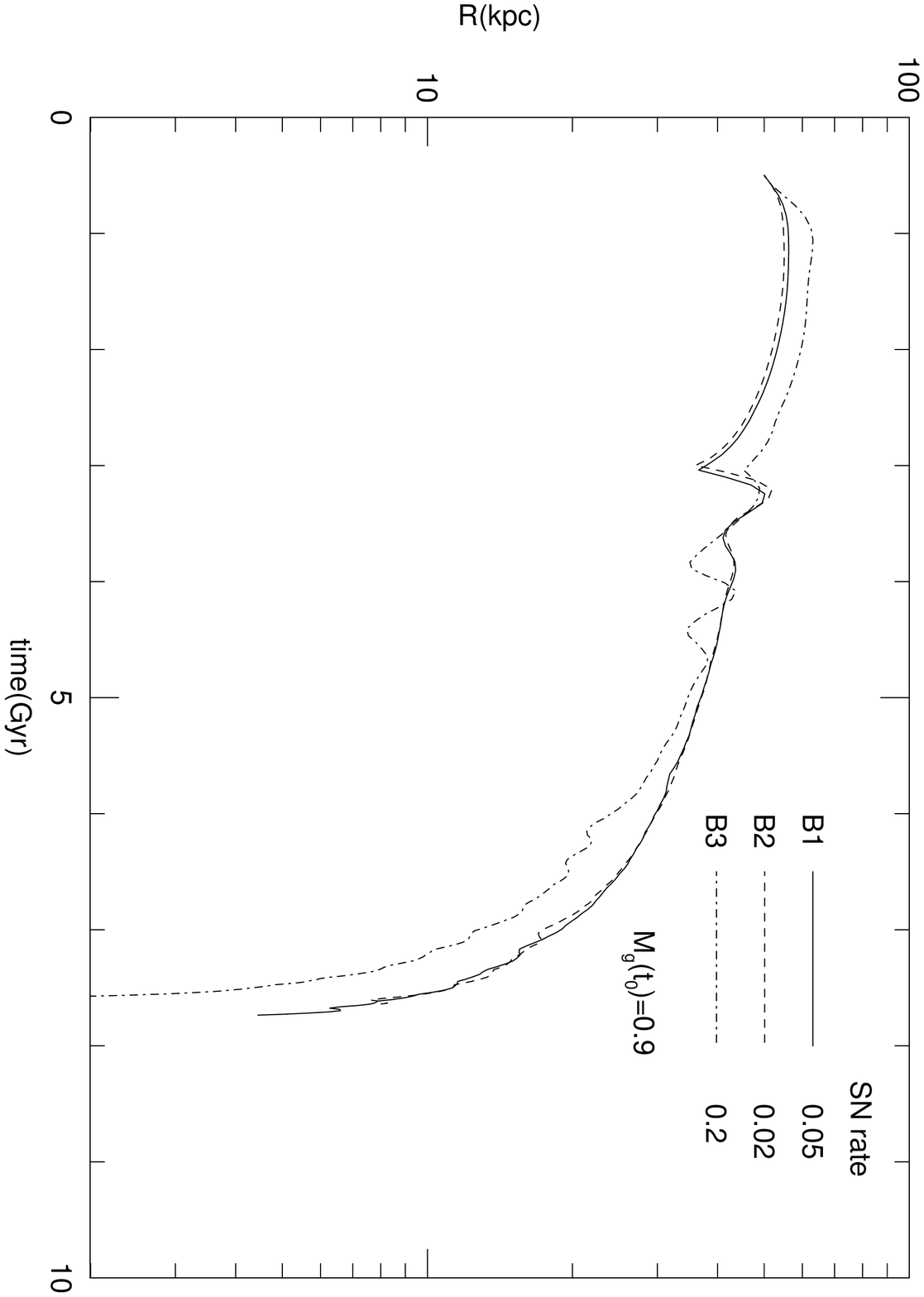]{The evolution of 
(a) radius (b) gas mass and (c) abundance for models B1 - B3.}

\figcaption[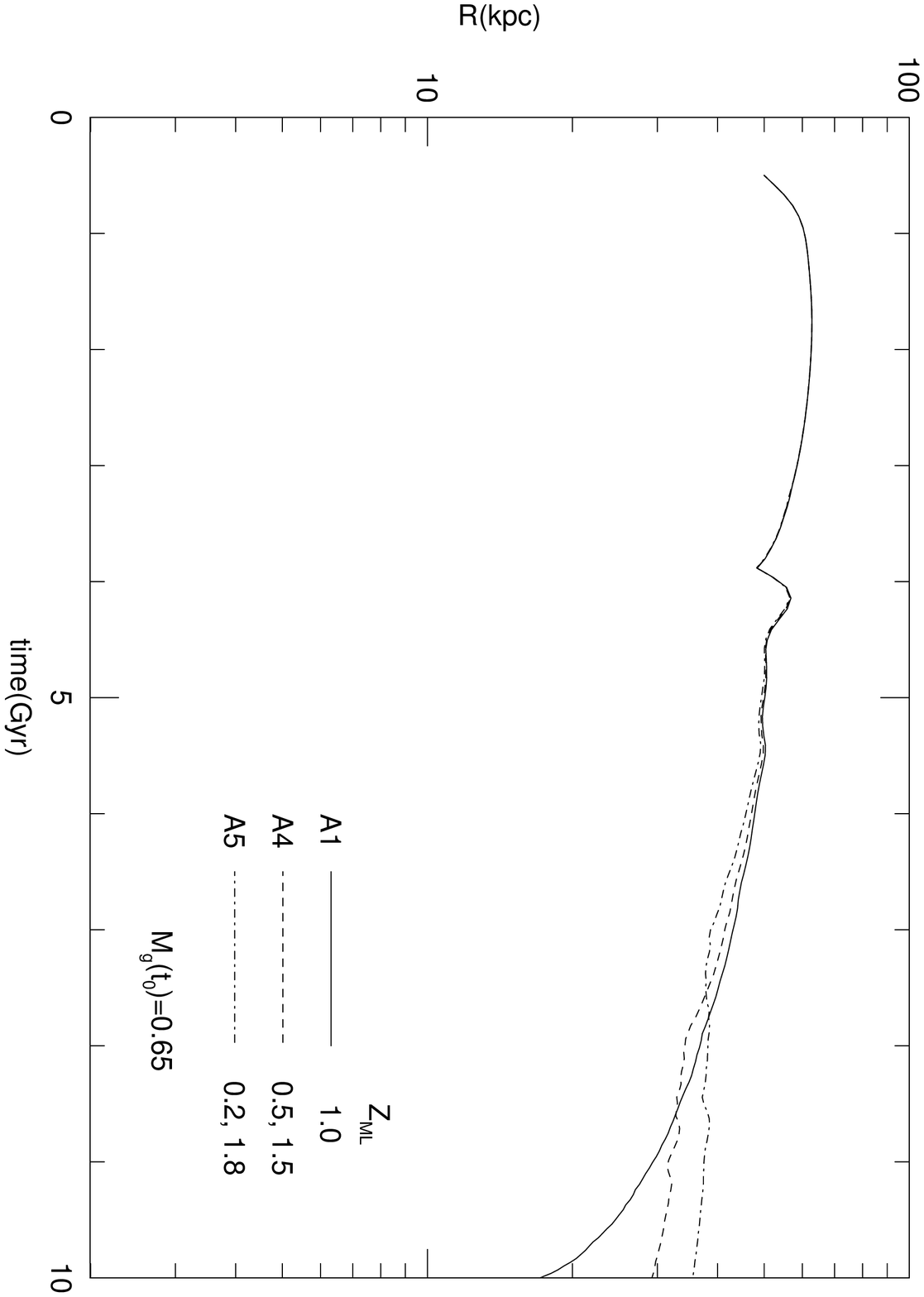]{The evolution of
(a) radius (b) gas mass and (c) abundance for models A1, A4 and A5.}

\figcaption[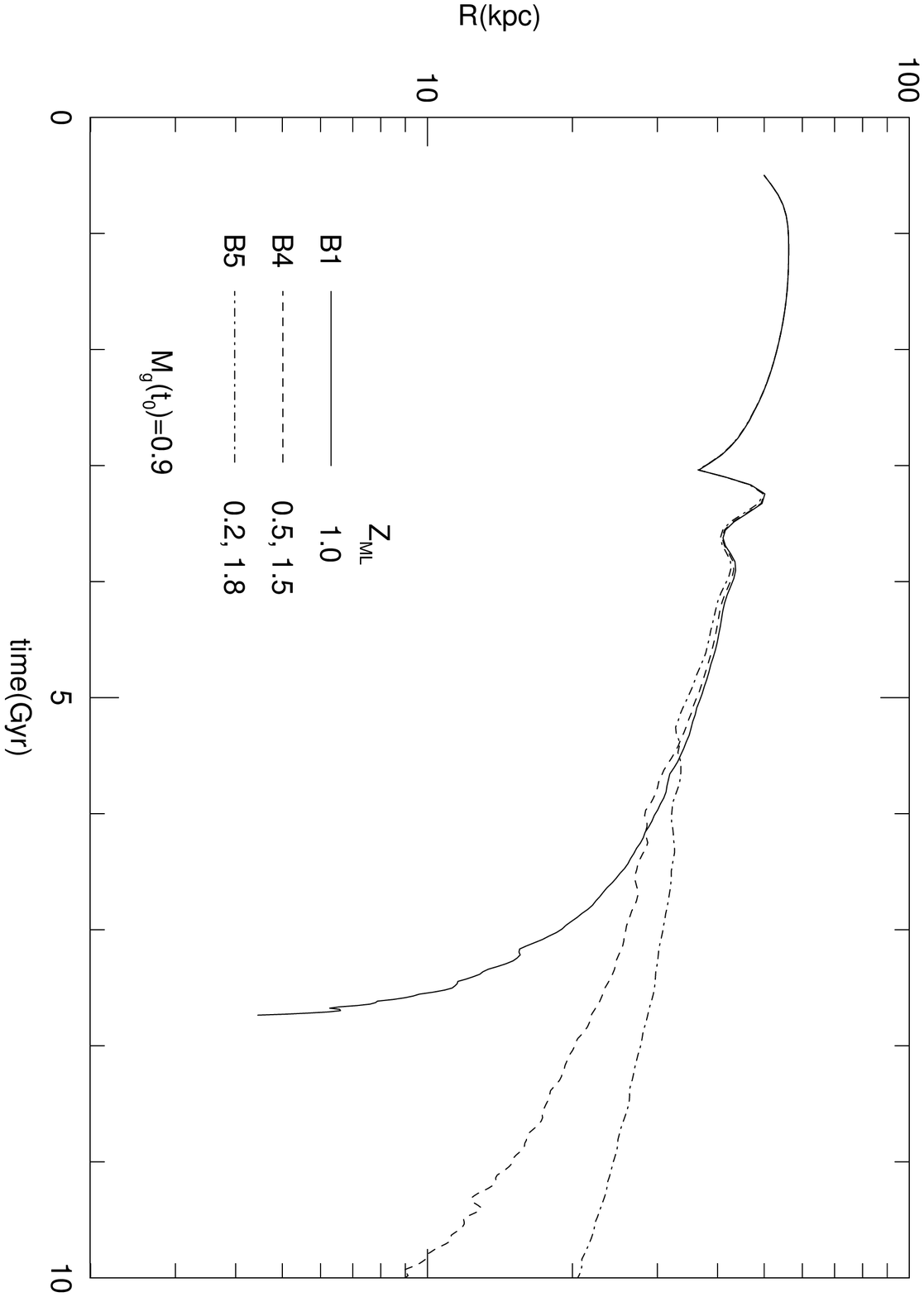]{The evolution of
(a) radius (b) gas mass and (c) abundance for models B1, B4 and B5.}

\figcaption[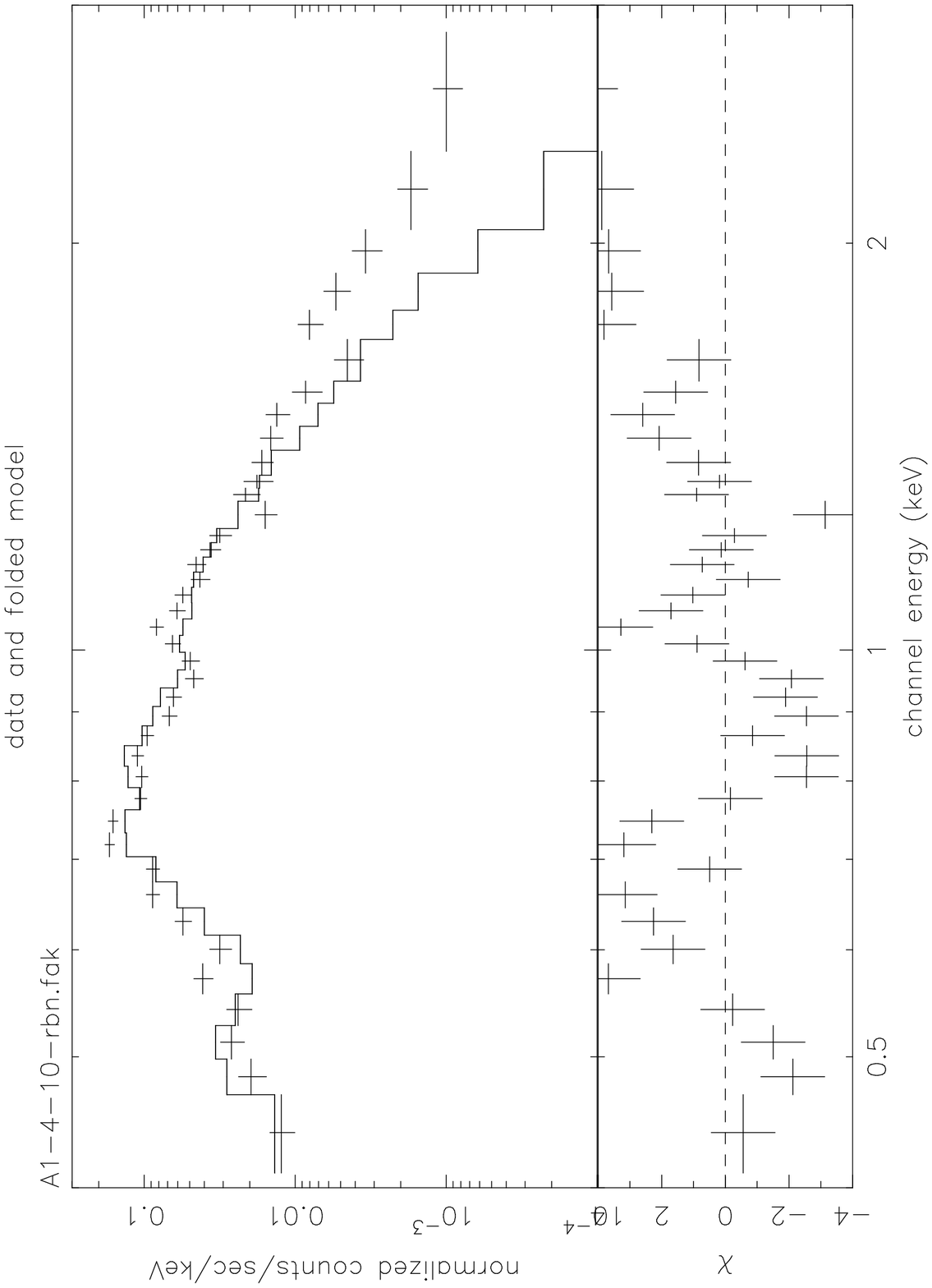]{Simulated X-ray spectra observed 
with the {\em ASCA} SIS. 
The line shows the best
fitting (one Meka plasma + absorption column density).
(a) Model A1 (b) Model A2 (c) Model B4 }

\figcaption[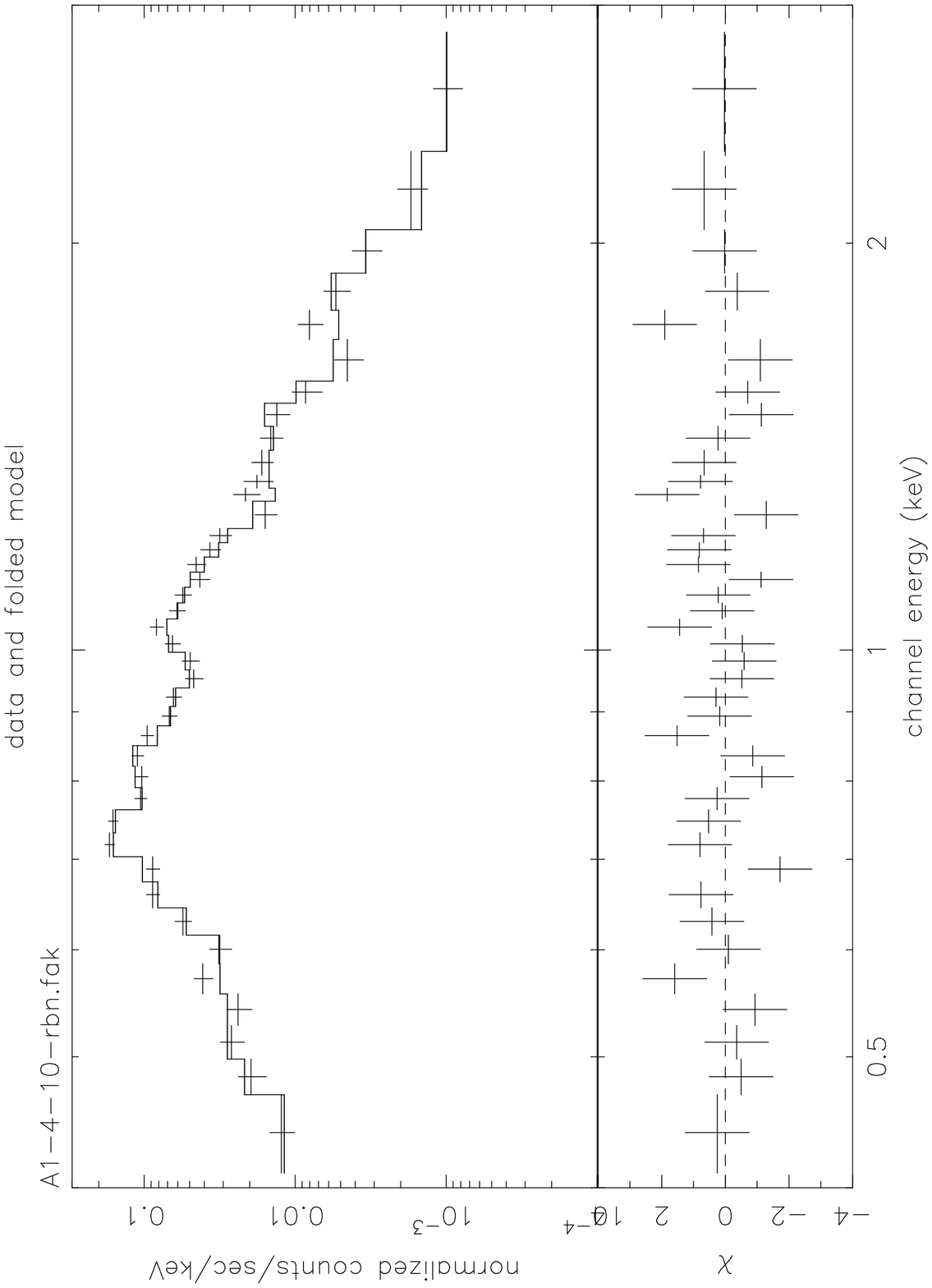]{Simulated X-ray spectra observed 
with the {\em ASCA} SIS. 
The line shows the best
fitting (two Meka plasma + absorption column density).
(a) Model A1 (b) Model A2 (c) Model B4 }


\begin{references}

\reference{ay1987} Arimoto, N., \& Yoshii, Y. 1987,
\aap, 173, 23

\reference{amt1994} 
Awaki, H., Mushotzky, R., Tsuru, T., Fabian, A. C., Fukazawa, Y.,
Loewenstein, M., Makishima, K., Matsumoto, H., Matsushita, K., Mihara,
T., Ohashi, T., Ricker, G. R., Serlemitsos, P. J., Tsusaka, Y., \&
Yamazaki, T. 1994,
\pasj,  46, L65

\reference{bt1987} Binney, J., \& Tremaine, S. 1987,
in Galactic Dynamics (Princeton : Princeton Univ. Press)

\reference{cdp1991} Ciotti, L., D'Ercole, A., Pellegrini, S., \& Renzini,
A. 1991,
\apj, 376, 380

\reference{cfn1980} Cowie, L. L., Fabian, A. C., \& Nulsen, P. E. J. 1980,
\mnras,  191, 399

\reference{dfj1990} David, L. P., Forman, W., \& Jones, C. 1990,
\apj, 359, 29

\reference{dfj1991} David, L. P., Forman, W., \& Jones, C. 1991,
\apj, 380, 39

\reference{fi1996} Fukumoto, J., \& Ikeuchi, S. 1996,
\pasj, 48, 1

\reference{i1997} Ikeuchi, S. 1977,
Prog. Theor. Phys., 58, 1742

\reference{k1992} Kaastra, J.S. 1992, An X-Ray Spectral Code for 
Optically Thin Plasmas
(Internal SRON-Leiden Report, updated version 2.0)

\reference{kf1995} Kim, D. -W., \& Fabbiano, G. 1995,
\apj,  441, 182

\reference{ka1991} K\"{o}ppen, J., \& Arimoto, N. 1991,
\aaps, 87, 109

\reference{l1974} Larson, R. B. 1974,
\mnras, 166, 585

\reference{mma1994} Matsushita, K., Makishima, K., Awaki, H., Canizares, C. R.,
Fabian, A. C., Fukazawa, Y., Loewenstein, M., Matsumoto, H., Mihara,
T., Mushotzky, R. F., Ohashi, T., Ricker, G. R., Serlemitsos, P. J.,
Tsuru, T., Tsusaka, Y., \& Yamazaki, T. 1994,
\apj,  436, L41

\reference{m1990} Mathews, W. G. 1990,
\apj, 354, 468

\reference{mg1995} Matteucci, F., \& Gibson, B. K. 1995, 
\aap, 304, 11

\reference{mt1987} Matteucci, F., \& Tornamb\'{e}, A. 1987,
\aap, 185, 51

\reference{mgv1985} 
Mewe, R., Gronenschild, E.H.B.M., \& van den Oord, G.H.J. 1985,
\aaps, 62, 197

\reference{mlf1986} Mewe, R., Lemen, J.R., \& van den Oord, G.H.J. 1986, 
\aaps, 65, 511

\reference{mt1994} Mihara, K., \& Takahara, F. 1994,
\pasj, 46, 447

\reference{rcdp1993} Renzini, A., Ciotti, L., D'Ercole, A., \&
Pellegrini, S. 1993,
\apj, 419, 52

\reference{s1978} Spitzer, L. 1978,
in Physical Processes in the Interstellare Medium (New York : Wiley)


\end{references}
\end{document}